\documentclass[a4paper]{article}

\usepackage{INTERSPEECH2020}

\usepackage{xcolor}
\newcommand\blfootnote[1]{%
	\begingroup
	\renewcommand\thefootnote{}\footnote{#1}%
	\addtocounter{footnote}{-1}%
	\endgroup
}

\title{High Quality Streaming Speech Synthesis with Low, Sentence-Length-Independent Latency}
\name{
Nikolaos Ellinas$^1$,
Georgios Vamvoukakis$^1$,
Konstantinos Markopoulos$^1$,
Aimilios Chalamandaris$^1$,
Georgia Maniati$^1$,
Panos Kakoulidis$^1$,
Spyros Raptis$^1$,
June Sig Sung$^2$,
Hyoungmin Park$^2$,
Pirros~Tsiakoulis$^1$
}
\address{$^1$Innoetics, Samsung Electronics, Greece\\
$^2$Mobile Communications Business, Samsung Electronics, Republic of Korea}
\email{\{n.ellinas, g.vamvouk, k.markop, aimilios.ch, g.maniati, p.kakoulidis, s.raptis, js6.sung, hm94.park, p.tsiakoulis\}@samsung.com}

\begin{document}

\maketitle
\begin{abstract}
This paper presents an end-to-end text-to-speech system with low latency on a CPU, suitable for real-time applications.
The system is composed of an autoregressive attention-based sequence-to-sequence acoustic model and the LPCNet vocoder for waveform generation.
An acoustic model architecture that adopts modules from both the Tacotron 1 and 2 models is proposed, while stability is ensured by using a recently proposed purely location-based attention mechanism, suitable for arbitrary sentence length generation.
During inference, the decoder is unrolled and acoustic feature generation is performed in a streaming manner, allowing for a nearly constant latency which is independent from the sentence length.
Experimental results show that the acoustic model can produce feature sequences with minimal latency about 31 times faster than real-time on a computer CPU and 6.5 times on a mobile CPU, enabling it to meet the conditions required for real-time applications on both devices.
The full end-to-end system can generate almost natural quality speech, which is verified by listening tests.
\end{abstract}

\section{Introduction}

\blfootnote{Proceedings of INTERSPEECH (doi: 10.21437/Interspeech.2020-2464)}

Developments in Deep Learning research and in computer hardware in the recent years have 
resulted in a paradigm shift in almost all speech processing applications.
For text-to-speech (TTS) in particular, there is a shift from either concatenative systems
or HMM-based statistical parametric models into neural models.
The latter have yielded very high quality synthetic speech
while at the same time the overall text-to-speech pipeline has been greatly simplified.

In most cases, the speech synthesis pipeline consists of an end-to-end acoustic model,
such as Tacotron \cite{wang2017tacotron} and a vocoder, such as WaveNet \cite{oord2016wavenet} or WaveRNN \cite{kalchbrenner2018efficient}.
Given a sequence of linguistic features the acoustic model predicts a sequence of speech frames in a parametric form (e.g. mel-spectrograms).
The sequence of acoustic features is then fed into the vocoder in order to synthesize the raw audio waveform at the desired sampling rate.
In the case of an end-to-end acoustic model the sequence of linguistic features consists of just the character sequence or the phoneme sequence.
Such models typically have a very high number of parameters resulting in high runtime complexity.
In spite of their impressive and realistic results, this
can be prohibitive for applications that require real-time speech synthesis on running devices without GPU,
e.g mobile phones, wearables and IoT devices.

\subsection{Related work}

Tacotron 2 \cite{shen2018natural} is the most popular acoustic model
that produces almost human-level speech when combined with the WaveNet vocoder.
It is an attention-based sequence-to-sequence model that leverages Recurrent Neural Networks (RNNs) and location sensitive attention \cite{chorowski2015attention}.
Escaping the necessity for RNNs which are considered complex and slow to train, convolutional models have also been proposed such as Deep Voice 3 \cite{ping2017deep} and DCTTS \cite{tachibana2018efficiently},
as well as a Transformer \cite{li2019neural} model which is based on the original architecture used for machine translation \cite{vaswani2017attention}.

Autoregressive models are characterized by slow inference speed, as the generation process is done sequentially.
There is always a model state, that has to be passed on to the next time-step for the model to be able to produce the acoustic frame.
ParaNet \cite{peng2019parallel}, ClariNet \cite{ping2018clarinet} and GAN-TTS \cite{Binkowski2020High} are fully convolutional models that attempt to escape the autoregressiveness of the above models and synthesize speech in parallel.
The latter two achieve direct text-to-audio synthesis, using knowledge distillation and adversarial training respectively.
FastSpeech \cite{ren2019fastspeech} is the most consistent attempt at solving the slow inference speed problem.
It consists of a Feed-Forward Transformer network that enables parallel acoustic frame generation, guided by a length regulator that produces the alignment between each linguistic unit and its corresponding number of acoustic frames.

As for vocoders, WaveNet \cite{oord2016wavenet} is the first autoregressive convolutional model that produces high quality audio when conditioned on mel-spectrograms.
It is also an example of very slow inference speed because of its sample-by-sample generation, which is later alleviated in Parallel WaveNet \cite{oord2017parallel} through parallel generation.
WaveGlow \cite{prenger2018waveglow} is a non-autoregressive model that uses normalizing flows and produces state-of-the-art results, but is slow for inference on CPU. 
Two more recent attempts that instead use adversarial training are MelGAN \cite{kumar2019melgan} and Parallel WaveGAN \cite{parallelwavegan}.

To our knowledge, the majority of previous research focuses on different architecture designs in order to produce the best speech quality possible, utilizing powerful GPUs both for training and inference.
There has not been given much attention to the performance of neural acoustic models when running on CPUs, which represent a more realistic scenario for some real-life applications.
In the field of vocoders, WaveRNN \cite{kalchbrenner2018efficient} and LPCNet \cite{valin2019lpcnet} are prime examples that a very simple architecture, in this case composed almost only from a couple of RNN cells, when designed properly can produce faster than real-time high quality results.

\subsection{Proposed method}

In this paper, we emphasize on optimizing the acoustic model, for real-life CPU applications.
We build on prior work from \cite{wang2017tacotron, shen2018natural, valin2019lpcnet, gmmattention} with the following contributions:
\begin{itemize}
\item A streaming inference method for purely autoregressive models with minimal latency that is independent of the sentence length
\item An overall lightweight end-to-end TTS architecture with low complexity that runs faster than real-time on CPU
\item The utilization of a robust alignment model that eliminates attention failure errors
\end{itemize}
We are particularly interested in a TTS system with state-of-the-art quality suitable for applications both on mobile devices and the server-side at low cost.
For such applications, there are variables that need to be considered when assessing the feasibility of a model besides its quality.
A significant hindrance for many models is that the inference time depends on the sentence length, which can lead to big slowdowns especially now that very long sentence generation is made possible through novel alignment models \cite{gmmattention}.
Another bottleneck is the latency from the beginning of the synthesis until audible speech is generated.

The proposed acoustic model combined with the streaming method can generate frames about 31 times faster than real-time on a single CPU thread, while simultaneously minimizing its latency down to about 50 ms.
On a mobile phone, these numbers change to 6.5 and 240 ms respectively because of the decreased capabilities of the mobile CPU.
The full synthesis including the waveform generation is done about 7 times faster than real-time on computer CPU and 2.7 times on mobile.

\section{Method}

\subsection{Acoustic model architecture}

The acoustic model maps the input sequence into a sequence of acoustic feature frames
that correspond to the representation used by the LPCNet vocoder.
It is an attention based sequence-to-sequence model and a direct modification of Tacotron 1 and 2 \cite{wang2017tacotron}\cite{shen2018natural}.

The encoder converts input sequences $\boldsymbol{p}=[p_1,...,p_N]$ to learnable embedding vectors, which are then processed by a 2-layer pre-net and a CBHG stack from \cite{wang2017tacotron} in order to produce the final encoder representation $\boldsymbol{e}=[e_1,...,e_N]$. On the decoder side, the inputs are acoustic frames $\boldsymbol{f}=[f_1,...,f_T]$ processed again by a pre-net. At each decoding step an attention RNN produces a hidden state $h_i$ by consuming the output of the previous step concatenated with the previous attention context vector. The attention module produces the current context vector, which is then fed to a stack of 2 residual decoder RNNs along with the attention RNN hidden state.

The output acoustic frames are predicted by a single fully-connected layer. When the decoding is complete, a residual is constructed from a 5-layer convolutional post-net from \cite{shen2018natural} and added to the output in order to increase the quality of the final outputs. Similar to Tacotron 2, a binary stop token that signals the end of the acoustic sequence is also predicted from a fully-connected layer with sigmoid activation.

\subsection{Alignment model}

In \cite{gmmattention}, a systematic comparison of various attention mechanisms is performed and shows that the purely location-based GMM attention introduced by Graves \cite{graves2013generating} is able to generalize to arbitrary sequence lengths and not violate the monotonicity of the learned alignment.
It does not rely on the encoder outputs for computing the scores, but instead has a state that is passed on to the next step making it the best candidate for our streaming model in order to be sentence-length-independent.

Our model uses a variation of GMM attention, similar to \cite{vasquez2019melnet}, where the mixture of Gaussian distributions is replaced by a Mixture of Logistic distributions (MoL), hence we refer to it as MoL attention. In order to compute the alignments, we directly use the Cumulative Distribution Function (\ref{cdf}) of the logistic distribution which is very simple to compute as it is equal to the sigmoid function.
\begin{equation}
	F(x;\mu,s)=\frac{1}{1+e^{-\frac{(x-\mu)}{s}}}=\sigma\left(\frac{x-\mu}{s}\right)
	\label{cdf}
\end{equation}
Hence, for each decoder step $i$ the alignment probabilities of each encoder timestep $j$ are computed as in (\ref{scores}) and the context vector as the weighted sum of the encoder representations (\ref{context}).
\begin{equation}
a_{ij} = \sum_{k=1}^{K}w_{ik}\left(F(j+0.5;\mu_{ik},s_{ik})-F(j-0.5;\mu_{ik},s_{ik})\right)
\label{scores}
\end{equation}
\begin{equation}
c_i = \sum_{j=1}^{N}a_{ij}e_j
\label{context}
\end{equation}
The parameters of the mixture are computed at each timestep as in equations (\ref{muik}-\ref{wik}) from the intermediate parameters $\hat{\mu}_{ik}$, $\hat{s}_{ik}$, $\hat{w}_{ik}$ which are predicted by 2 fully connected layers (\ref{mlp}) applied to the attention RNN state $h_i$.
\begin{equation}
\mu_{ik}=\mu_{i-1k}+\exp (\hat{\mu}_{ik})
\label{muik}
\end{equation}
\begin{equation}
s_{ik}=\exp (\hat{s}_{ik})
\label{sik}
\end{equation}
\begin{equation}
w_{ik}=softmax(\hat{w}_{ik})
\label{wik}
\end{equation}
\begin{equation}
\left(\hat{\mu}_{ik},\hat{s}_{ik},\hat{w}_{ik}\right)=W_2\tanh(W_1(h_i))
\label{mlp}
\end{equation}

\subsection{Vocoder}

We use the LPCNet \cite{valin2019lpcnet} vocoder as adapted for reduced complexity by the parallel work of \cite{srlpcnet}.
By providing an initial estimation of the spectral envelope through Linear Prediction Coefficients (LPC), this method allows the neural model to focus on modeling the excitation signal, thus enabling it to produce state-of-the-art speech quality with a significantly lower complexity. The conditioning parameters required for LPCNet are used as a target for our acoustic model, so that we have an end-to-end speech synthesis pipeline. 

\subsection{Streaming inference}

\begin{figure*}
  \includegraphics[width=\linewidth]{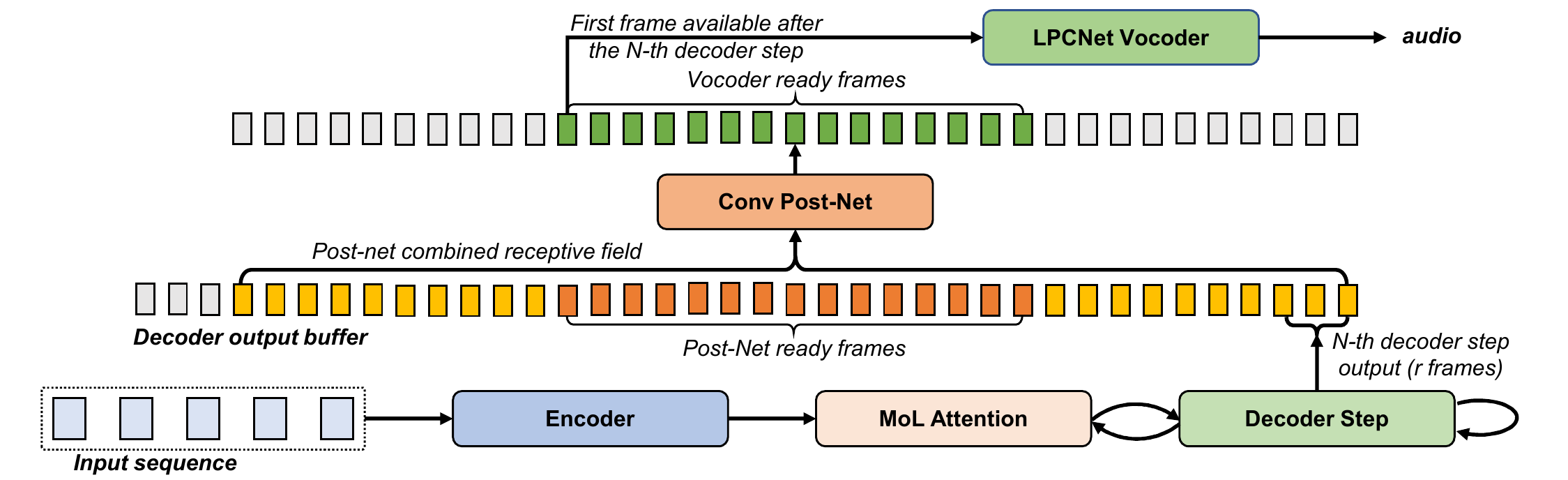}
  \caption{Streaming inference architecture.
  The input sequence is fed into the encoder, which produces the linguistic representations.
  The decoder, conditioned on the context vector produced by the attention module, generates a small batch of acoustic frames at each step which are buffered.
  When the buffer contains enough frames, they are passed to the post-net in a larger batch.
  Care is taken so that the buffer includes enough frames to accommodate the total
  receptive field of the convolutional layers in the post-net.
  Finally, the output acoustic frames are passed to the LPCNet vocoder,
  after discarding the initial frames that were already vocoded but were included as part of the receptive field
  and the most recent frames that need additional future frames to be processed by the post-net.}
  \label{fig:architecture2}
  \vspace{-0.6em}
\end{figure*}

During speech generation, the acoustic frame sequence must be produced and then fed into the vocoder.
A model like FastSpeech \cite{ren2019fastspeech} can generate the whole sequence and then feed it to a vocoder like WaveGlow \cite{prenger2018waveglow} which in turn can generate the full raw speech utterance.
This process can take very little time on GPU due to the memory efficient parallel computations.
However, for longer utterances that need to be synthesized on CPU, this process can take a much longer time, thus increasing the latency from when the synthesis starts until the final audible speech is generated.

We implement a streaming inference process that enables the feeding of acoustic frames into the vocoder before the inference process of the acoustic model is finished
as shown in Figure~\ref{fig:architecture2}.
The LPCNet vocoder is autoregressive, so it can start generating after receiving the first acoustic frame and because it runs faster than real-time, the resulting audio can be audible almost immediately.

In order to minimize latency, we continuously gather acoustic frames in groups that are passed in parallel to the post-net
taking advantage of its convolutional nature.
The post-net is trained to refine the full acoustic frame sequence, but since it is fully convolutional it can perform the same process in a smaller sequence just as well.
The output frames from each decoder step are accumulated in a buffer and then sent in larger chunks to the post-net.
A small window of frames before and after the frame segments to be synthesized should be kept and fed to the post-net
as determined by the total receptive field of the convolutional layers (21 in our model),
in order for the output to be identical to the non-streaming version of the model.
The number of frames in each chunk sent to the post-net is a trade-off between the latency and the real-time-factor (RTF) of the system.
If it is small then the latency is small, as the first frames available for vocoding require a few decoder steps,
but the computational overhead of the window frames is big hindering the real-time-factor. 
On the other hand if the number of frames accumulated for post-net processing is too big 
the window frames overhead diminishes but the latency increases as we need to wait for more decoder steps.
A value of 100 frames is chosen for the chunk sent to the post-net that corresponds to 1 second of speech,
as it is common practice to buffer some amount of audio before playback in order to avoid static, especially if the audio is transmitted over the network.
Then the vocoder synthesizes the first 1 second of audio from the output of the post-net.
With the first second of audio synthesized, the user can start listening while the next audio segment is being generated.
A detailed visualization of the proposed method can be seen in Figure~\ref{fig:architecture2}.

The only requirement for this method to be feasible in a production environment, is that the CPU can run the process faster than real time. 
This way a low and stable latency is guaranteed, regardless of sentence length. 
In our experiments, we found that a CBHG-based post-net architecture is also feasible and although the final output is not identical to the non-streaming version due to the bidirectional GRU layer,
the quality is not significantly affected if the window is adjusted properly.

\section{Experiments and results}

\subsection{Experimental setup}

We train our model on the LJ Speech dataset \cite{ljspeech17}, after upsampling the audio data to 24 kHz.
The acoustic features used for training are matching the ones by the LPCNet vocoder \cite{valin2019lpcnet}, i.e. 20 Bark-scale cepstral coefficients (increased by 2 bands compared to LPCNet because of the higher sampling rate), the pitch period and pitch correlation.

The input text is first normalized and converted into a phoneme sequence by a traditional TTS front-end module,
though without any modification the proposed method can be applied to character-based models.
In the encoder, phonemes are mapped into 256 dimensional embeddings and the GRU of the CBHG module has 128 dimensions in each direction.
The decoder contains 3 RNNs, a 256-dimensional attention GRU and two 512-dimensional residual decoder LSTMs.
The attention module uses a mixture of 5 logistic distributions and its feed-forward layers are 256-dimensional.
Pre-net and post-net layers are regularized by dropout \cite{srivastava2014dropout} of rate 0.5 and the decoder LSTMs by zoneout \cite{zoneout} of rate 0.1.

The network parameters are trained with the Adam optimizer \cite{adam}, which minimizes the average L1 loss before and after the post-net, batch size 32 and an initial learning rate of $10^{-3}$ linearly decaying to $3\cdot10^{-5}$ after 100,000 iterations. L2 regularization with weight $10^{-6}$ is also used.
For our implementation, we use the PyTorch framework \cite{paszke2017automatic}.

\subsection{Complexity}

The decorrelation of the acoustic information into cepstrum and pitch and the very low dimensionality of the used features, allows us to reduce the model parameters without affecting output quality.
The proposed model as described in the previous section has a total of 9.5 million parameters.

We also take advantage of the model's ability to generate batches of frames at each decoder time step instead of a single frame.
By increasing the number of frames per step ($r$) the RTF can be reduced, as the decoder runs for fewer steps given the same linguistic input, as shown in Table~\ref{tab:rtf}. However, the quality is also dependent on $r$, so we adopt the value $r=5$ as we find it provides a good trade-off between speed and quality.

Inference is done on a single thread on an i7-8700K CPU at 3.7 GHz, as we are targeting a realistic server scenario where there are multiple TTS requests running in parallel on the same server.
We also report that for $r=5$ on a mobile Exynos 9820 CPU the RTF is 0.153 $\pm$ 0.006 for the acoustic model, while when including the vocoder, the RTF of the total system is 0.136 $\pm$ 0.005 ms on computer CPU and 0.372 $\pm$ 0.007 ms on mobile.

\begin{table}[h]
	\caption{Average RTF and latency (with standard deviation) for different values of the number of frames per step ($r$) and corresponding MOS (with 95\% confidence interval) on the test corpus.
	The RTF and latency measurements are for the acoustic model inference only, without the vocoder inference time.}
	\label{tab:rtf}
	\centering
	\begin{tabular}{lccc}
		\toprule
		$r$ & RTF       & Latency (ms)       & MOS     \\
		\midrule
		2  & 0.067 $\pm$ 0.005 &  93.3 $\pm$ 10.9 & 4.00 $\pm$ 0.13  \\
		3  & 0.044 $\pm$ 0.002 &  70.4 $\pm$ 9.1  & 4.23 $\pm$ 0.14  \\
		5  & 0.032 $\pm$ 0.002 &  50.1 $\pm$ 7.7  & 4.20 $\pm$ 0.10  \\
		7  & 0.025 $\pm$ 0.001 &  40.8 $\pm$ 7.0  & 4.01 $\pm$ 0.16  \\
		10 & 0.020 $\pm$ 0.001 &  34.1 $\pm$ 6.7  & 1.63 $\pm$ 0.13  \\
		\midrule
		\vspace{0.1em}
		   &                   &        Ground truth          & 4.5 $\pm$ 0.10 \\
		\cline{3-4}
	\end{tabular}
\vspace{-1.2em}
\end{table}

\subsection{Latency}

In order to measure the effectiveness of our method, we compute the latency in both streaming and non-streaming inference setups and also compare the results with other approaches.
For comparison we selected the ESPnet toolkit \cite{hayashi2019espnettts} because it provides many state-of-the-art pretrained models.
We tested the Tacotron 2 \cite{shen2018natural}, Transformer \cite{li2019neural} and FastSpeech \cite{ren2019fastspeech} models which utilized all available CPU threads for parallelization of operations, as per their original implementation.
We also measured the latency of the fastest model, FastSpeech, running on a single thread for a better comparison with our model.
Note that these models were trained on audio data at 22.05 kHz sampling rate and with 11.6 ms frame shift, while we use 24 kHz sampling rate and 10 ms frame shift in order to match the vocoder requirements.
As a result, our model needs to produce more frames for the same duration of audio.

\begin{figure}[t]
	\centering
	\includegraphics[width=\linewidth]{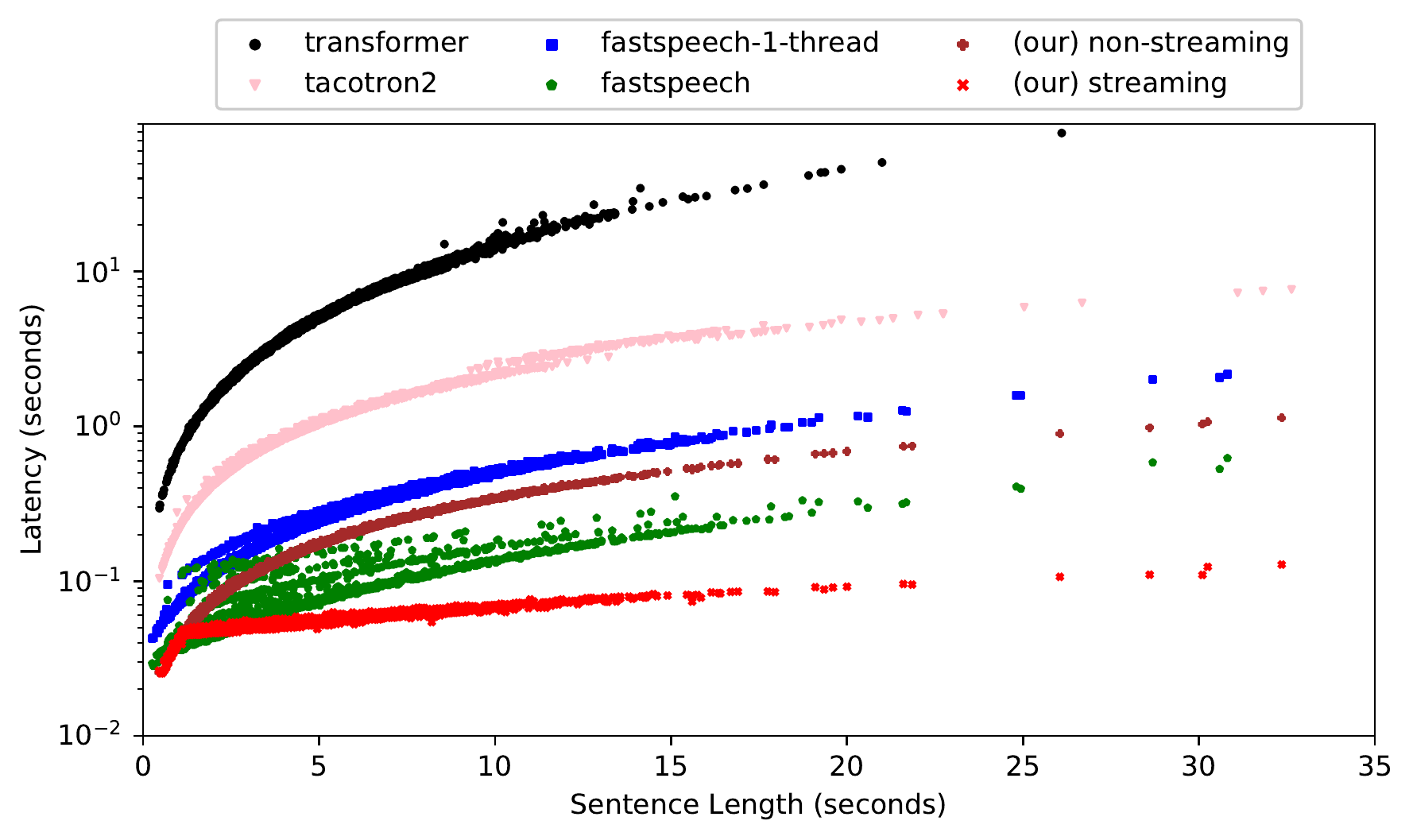}
	\caption{Latency vs sentence length. A logarithmic scale is used for the latency (y-axis) 
	in order to better visualize the large discrepancies between the various systems shown.
	The time duration is used to measure sentence length (x-axis) since not all systems have the same frame shift and front-end processing.
	}\vspace{-0.8em}
	\label{fig:times}
\end{figure}

In these experiments, we are interested in the latency of the acoustic model, i.e. the time interval measured from the beginning of the synthesis until the desired number of frames is generated.
A test corpus of 2213 sentences with various lengths was used in order to have accurate measurements of how the sentence length correlates to the latency and the results are shown in Figure~\ref{fig:times}.
Average latency values are also included in Table~\ref{tab:rtf} and show its dependence on the $r$ parameter.
On mobile CPU, the value corresponding to $r=5$ is 240 ms.

By examining Figure~\ref{fig:times}, we notice that the latency for all non-streaming inference models increases as the sentence length increases.
It is clear that our non-streaming model follows a similar pattern, but is faster due to its lightweight architecture.
FastSpeech is faster on multi-threaded run mode, but it becomes slower than our non-streaming model when running on a single thread.
On the other hand, the streaming model has an almost flat curve showing that its latency increases very slowly as a function of the sentence length.
The slight increasing tendency is exclusively attributed to the runtime of the encoder module that being autoregressive is dependent on the input sequence length.
However, the percentage of the encoder complexity is minimal, especially if we also consider the complexity of the vocoder, making the overall latency of the system practically sentence-length-independent.


\subsection{Quality}

We ran a listening test to investigate how quality changes by varying the $r$ parameter. 
We randomly selected 40 sentences from the benchmark corpus and synthesized them with each system trained on the LJ Speech dataset\footnote{Samples available at https://innoetics.github.io/}.
The listeners were presented a web-page containing shuffled audio samples and were asked to score the naturalness of each sample on a 5-point Likert scale.
The Mean Opinion Score (MOS) for each system including the ground truth is shown in Table~\ref{tab:rtf}.
For the larger $r$ values 7 and 10, we notice quality degradation, which is attributed to the inability of the model to adequately represent the longer speech information in a single step.
Especially for $r=10$, the model skips some of the phonemes in the input sequence, leading to completely unnatural speech.
For $r$ values 3 and 5, the quality is similar while for $r=2$ the MOS is lower as the model may have not fully converged due to its reduced complexity.

The subjective evaluation shows that our model apart from being very fast, also scores very high in terms of naturalness.
At the same time, we found no errors due to failed alignment in the benchmark corpus meaning that the synthesis is very robust thanks to the MoL attention module.
\vspace{-0.4em}
\section{Conclusions}

We have presented an autoregressive TTS acoustic model that can produce high quality speech many times faster than real time.
The combination of the lightweight model architecture with the proposed streaming inference method is ideal for real-time applications on both large scale systems and smaller devices as it offers a low and stable latency without the need of expensive hardware like GPUs.
The streaming generation outperforms other tested TTS systems in terms of latency and does not hurt the output quality, because the receptive field of the convolutional post-net is taken into consideration. 
The attention mechanism that was selected provides great stability even in very long sentences, which is an absolute requirement in real applications.
Further work can be made with improvements on runtime by taking advantage of sparsification and quantization methods for better performance which can enable the system to be run on even more low-tier devices.


\bibliographystyle{IEEEtran}

\bibliography{mybib}


\end{document}